\documentclass[aps,preprint,preprintnumbers,amsmath,amssymb,nofootinbib]{revtex4}
\usepackage{amssymb}
\usepackage{amsmath}
\usepackage{bm}

\begin{document}
\title{Stochastic emergence of inflaton fluctuations in a SdS primordial universe with large-scale repulsive gravity from a 5D vacuum.}
\author{$^{1}$ L. M. Reyes \footnote{
E-mail address: luzreyes@fisica.ugto.mx}, $^{1}$ Jos\'{e} Edgar Madriz Aguilar\footnote{
E-mail address: jemadriz@fisica.ugto.mx}, and  $^{2,3}$ Mauricio
Bellini \footnote{ E-mail address: mbellini@mdp.edu.ar,
mbellini@conicet.gov.ar} }
\address{$^{1}$ Departamento de F\'{\i}sica, DCI, Universidad de Guanajuato, Lomas del Bosque 103, Col. Lomas del Campestre,
C.P. 37150 Le\'on Guanajuato,
M\'exico.\\
$^2$ Departamento de F\'isica, Facultad de Ciencias
Exactas y Naturales, Universidad Nacional de Mar del Plata, Funes
3350,
C.P. 7600, Mar del Plata, Argentina.\\
$^3$ Instituto de Investigaciones F\'{\i}sicas de Mar del Plata (IFIMAR), \\
Consejo acional de Investigaciones Cient\'ificas y T\'ecnicas
(CONICET), Argentina.}

\begin{abstract}
We develop a stochastic approach to study scalar field
fluctuations of the inflaton field in an early inflationary
universe with a black-hole (BH), which is described by an
effective 4D Schwarzschild-de Sitter (SdS) metric. This effective 4D metric is the induced metric on a 4D hypersurface, here representing our universe, which is obtained from a 5D Ricci-flat SdS static metric, after implement a planar coordinate transformation. On this background we found that at
the end of inflation, the squared fluctuations of the inflaton
field are not exactly scale independent and result sensitive to
the mass of the BH.
\end{abstract}
 \maketitle

\section{Introduction}

In stochastic inflation the dynamics of the inflaton field is described by a second order stochastic
equation, where the emergence of a long-wave classical field that
drives inflation is subject to a short-wave classical noise.
Starobinsky \cite{sta} has noted that under certain assumptions,
the splitting of the scalar field into long-wavelength and
short-wavelength components leads to a quantum Langevin equation
that could become classical stochastic dynamics for the
long-wavelength modes of the scalar field. This approach
emphasizes the role of the quantum fluctuations as the driving
forces of the inflation. It considers as a main ingredient the set
of long-wavelength modes as a whole, from which the coarse-grained
field emerges. This coarse-grained field is assumed to have a
highly classical behavior, but the inflow of short-wavelength
modes alters its evolution in a random way. Furthermore, the
quantum fluctuations of the short-wavelength field, give place to cosmological density
perturbations that could be the origin of the structure of the
universe \cite{guth}. The coarse-graining
representation of the inflaton field has played an important role
in 4D standard inflationary
cosmology \cite{bcms}, in 5D inflationary cosmology from modern
Kaluza-Klein theory \cite{5D}, and in extensions to vectorial fields
more recently implemented in the framework of
Gravitoelectromagnetic Inflation \cite{gi}.

On the other hand, in the last years theories with extra dimensions have become quite popular in the scientific
community \cite{rs}. In particular, some brane scenarios \cite{br}, and
the induced matter (IM) theory of gravity \cite{im}, have been subject of
a great amount of research. Even when both
theories have different physical motivations for the introduction
of a large extra dimension, they are equivalent each
other, and predict identical non-local and local high
energy corrections to general relativity in 4D, and usual matter
in 4D is a consequence of the metric dependence on the fifth extra
coordinate \cite{pdl}.

On the basis of the IM theory, we have recently shown in \cite{mb} that there exists a 5D SdS BH solution of the theory, from which we can derive a 4D cosmological model where gravity manifests itself as attractive on small (planetary
and astrophysical) scales, but repulsive on very large
(cosmological) scales. This behavior of gravity derived from this 5D framework, leave us to put on the desk
the following question: can repulsive gravity be considered as a
strong candidate for explaining the large-scale accelerated
expansion of the universe in the past and today? To answer this
question let us to start by defining the physical vacuum via the 5D Ricci-flat metric \cite{antigravity}:
\begin{equation}\label{a1}
dS_{5}^{2}=\left(\frac{\psi}{\psi_0}\right)^{2}\left[c^{2}f(R)dT^{2}-\frac{dR^2}{f(R)}-R^{2}(d\theta^{2}+sin^{2}
\theta d\phi^{2})\right]-d\psi^{2}.
\end{equation}
Here, $f(R)=1-(2G\zeta\psi_{0}/Rc^2)-(R/\psi_0)^{2}$ is a
dimensionless function, $\lbrace T,R,\theta,\phi\rbrace$ are the
usual local spacetime spherical coordinates employed in general
relativity and $\psi$ is the non-compact space-like extra dimension.
In this line element $\psi_0$ is an arbitrary constant
with length units, $c$ denotes the speed of light, and the constant parameter $\zeta$ has units of
$(mass)(length)^{-1}$. This static metric is a 5D extension of the 4D SdS metric. In order to get this metric written on
a dynamical chart coordinate $\lbrace t,r,\theta,\phi \rbrace$, we use the coordinate transformation given by \cite{planar}
\begin{equation}\label{a2}
R=ar\left[1+\frac{G\zeta\psi_0}{2ar}\right]^{2},\quad
T=t+H\int^{r}dR\,\frac{R}{f(R)}\left(1-\frac{2G\zeta\psi_0}{R}\right)^{-1/2},
\psi=\psi,
\end{equation}
$a(t)=e^{Ht}$ being the scale factor, and $H$ the Hubble constant.
Thus the line element (\ref{a1}) can be written in terms of the
conformal time $\tau$ as
\begin{equation}\label{a3}
dS_{5}^{2}=\left(\frac{\psi}{\psi_0}\right)^{2}\left[F(\tau,r)d\tau^{2}-J(\tau,r)\left(dr^{2}+r^{2}(d\theta^{2}
+sin^{2}\theta d\phi^{2})\right)\right]-d\psi^{2},
\end{equation}
where the metric functions $F(\tau,r)$ and $J(\tau,r)$ are given by
\begin{equation}\label{a4}
F(\tau,r)=a^{2}(\tau)\left[1-\frac{G\zeta\psi_0}{2a(\tau)r}\right]^{2}\left[1+\frac{G\zeta\psi_0}{2a(\tau)r}\right]^{-2},\quad J(\tau,r)
=a^{2}(\tau)\left[1+\frac{G\zeta\psi_0}{2a(\tau)r}\right]^{4},
\end{equation}
with $d\tau=a^{-1}(\tau)dt$ and $a(\tau)=-1/(H\tau)$, so that the
Hubble parameter is a constant given by $H=a^{-2}\, {d a\over
d\tau}$. As it was shown in \cite{antigravity}, for certain values
of $\zeta$ and $\psi_0$, both metrics (\ref{a1}) and (\ref{a3}) have two natural
horizons. The inner horizon is the analogous of the Schwarzschild
horizon and the external one is the analogous of the Hubble
horizon.

Now we consider a 5D massless scalar field which is free of any
interactions: $^{(5)}\Box \varphi=0$.  We assume that
$\varphi(\tau,r,\theta,\phi,\psi)$ can be separated in the form
$\varphi(\tau,r,\theta,\phi,\psi)\sim
\Phi(\tau,r)G(\theta,\phi)\Omega(\psi)$, so that the expression
$^{(5)}\Box \varphi=0$ leaves to
\begin{eqnarray}
&&\left(\frac{\psi}{\psi_0}\right)^{-2}\frac{d}{d\psi}\left[\left(\frac{\psi}{\psi_0}\right)^{4}\frac{d\Omega}{d\psi}\right]+M^2\Omega = 0, \label{a9a}\\
\label{a9} && \frac{1}{\sqrt{FJ}}\frac{\partial}{\partial
\tau}\left(\sqrt{\frac{J^3}{F}}\frac{\partial\Phi}{\partial\tau}\right)-\frac{1}{2}
\left(\frac{1}{F}\frac{\partial F}{\partial
r}+\frac{1}{J}\frac{\partial J}{\partial
r}\right)\frac{\partial\Phi}{\partial r}
-\frac{1}{r^2}\frac{\partial}{\partial
r}\left(r^{2}\frac{\partial\Phi}{\partial r}\right) \nonumber
\\
&&-\left(\frac{l(l+1)}{r^2}-M^2J\right)\Phi=0 \label{a10},
\end{eqnarray}
where $M^2 >0$ is a separation constant with mass units and $l$ is
an integer dimensionless parameter related with the angular
momentum.

\section{The dynamics of $\varphi$ on the 4D hypersurface $\Sigma$}

Assuming that the 5D spacetime can be foliated by a family of
hypersurfaces $\Sigma:\psi=\psi_0$, from the metric (\ref{a3}) we
obtained that the 4D induced metric on every leaf $\Sigma$ is
given by
\begin{equation}\label{b1}
dS_{4}^{2}=F(\tau,r)d\tau^{2}-J(\tau,r)[dr^{2}+r^{2}(d\theta^{2}+\sin^{2}\theta d\phi^{2})],
\end{equation}
where the metric functions $F(\tau,r)$ and $J(\tau,r)$ can be now written in terms of the physical mass $m=\zeta\psi_0$ (introduced by the first
time in \cite{antigravity}), in the form
\begin{equation}\label{b2}
F(\tau,r)=a^{2}(\tau)\left[1-\frac{Gm}{2a(\tau)\,r}\right]^{2}\left[1+\frac{Gm}{2a(\tau)\,r}\right]^{-2},\quad
J(\tau,r)=a^{2}(\tau)\left[1+\frac{Gm}{2a(\tau)\,r}\right]^{4}.
\end{equation}
The induced metric (\ref{b1}) has a Ricci scalar $^{(4)} {\cal R}
= 12 H^2$. describes a black hole in an expading universe, where
the expansion is driven by a kind of cosmological constant, whose
value in general depends of the value of $\psi_0$. According to \cite{antigravity}, this metric also indicates that there exists a  length scale that separates regions on which gravity changes from
attractive to repulsive. This length scale is called the gravitational-antigravitational
radius, which in the coordinates $(T,R)$ is given by
$R_{ga}=(Gm\psi^2_0)^{1/3}$. With the help of
(\ref{a2}), in the new coordinates $(\tau,r)$ this radius must
obey the relation
\begin{equation}\label{b6}
r_{ga}=\frac{1}{2a(\tau)}\left[R_{ga}-Gm\pm\sqrt{R_{ga}^{2}-2GmR_{ga}}\right],
\end{equation}
where $r_{ga}$ is denoting the gravitational-antigravitational
radius in the new coordinates and the solution with the minus sign is not physical.
In order to $r_{ga}$ to be a real value quantity, we
require the condition $R_{ga}^{2}-2GmR_{ga}\geq 0$ to be hold.
This condition can be rewritten in the form $m^{2}\leq\frac{\psi_{0}^{2}}{8G^{2}}$. Thus, if we consider the foliation $\psi_0=c^2/H$ and the fact that for
$c=\hbar=1$ the Newtonian constant is
$G=M^{-2}_p$\footnote{$M_p=1.2 \times 10^{19}\,$ GeV is the
Planckian mass.}, this condition leaves now to the restriction $\epsilon=\frac{ m H}{M^2_p} \le \frac{1}{2 \sqrt{2}} \simeq
0.353553$. This way, the model allows to consider objects whose mass sa\-tis\-fies
the parameter condition: $\epsilon = GmH \ll 1$. For these values, $R_{ga}$ is smaller than the size of our universe
horizon. The same restriction has been used in \cite{CNW} with different motivation.\\

Now, from (\ref{a9a}) and (\ref{a10}), the 4D induced field equation reads
\begin{equation}\label{b3}
\frac{1}{\sqrt{FJ^3}}\frac{\partial}{\partial\tau}\left[\sqrt{\frac{J^3}{F}}\frac{\partial\,{\bar\varphi}}{\partial\tau}\right]-\frac{1}{2}
\left(\frac{1}{FJ}\frac{\partial F}{\partial
r}+\frac{1}{J^2}\frac{\partial J}{\partial
r}\right)\frac{\partial\,{\bar\varphi}}{\partial r}
-\frac{1}{J}\nabla^{2}\,\bar\varphi + M^2\,{\bar\varphi}=0,
\end{equation}
where ${\bar\varphi}(\tau,r,\theta,\phi)=\varphi(\tau,r,\theta,\phi,\psi_0)$ is the effective scalar field induced on the generic hypersurface $\Sigma$, which we shall identify with the inflaton field. It can be easily seen from (\ref{b3}) that $M$ here corresponds to the physical mass of $\bar{\varphi}$. We can expand the field ${\bar\varphi}$ as
\begin{equation}
\bar\varphi(\vec r,\tau) = \int^{\infty}_{0} dk\,\sum_{lm}
\left[a_{klm} \bar\Phi_{klm}(\vec r,\tau)+a^{\dagger}_{klm}
\bar\Phi^*_{klm}(\vec r,\tau)\right],
\end{equation}
where $\bar\Phi_{klm}(\vec r,\tau)= k^2 \,j_l\left(kr\right)
\bar\Phi_{kl}(\tau) Y_{lm}(\theta,\phi)$, $Y_{lm}(\theta,\phi)$
are the spherical harmonics, $j_{l}(kr)$ are the spherical Bessel
functions and the annihilation and creation
operators obey: $\left[a_{klm}, a^{\dagger}_{k'l'm'}\right] = \delta(k- k')
\delta_{ll'} \delta_{m m'}$, $ \left[a_{klm},
a_{k'l'm'}\right]=\left[a^{\dagger}_{klm},
a^{\dagger}_{k'l'm'}\right]=0$. Hence, using the addition theorem for spherical harmonics, we obtain for the mean squared fluctuations
\begin{equation}
\left< 0\left| \bar\varphi^2\left(\vec r,\tau\right)
\right|0\right> = \int^{\infty}_{0} \frac{dk}{k} \sum_{l}
\frac{2l+1}{4\pi} k^5 j^2_l(kr)
\left|\bar\Phi_{kl}(\tau)\right|^2. \label{fluct}
\end{equation}
Now, if we assume that
$\bar{\varphi}_l(\tau,r,\theta,\phi)=\bar{\Phi}_l(\tau,r)\bar{G}_{l,m}(\theta,\phi)$,
then the equation for $\bar\Phi_l(r,\tau)$ on the hypersurface
$\Sigma$ can be written as
\begin{eqnarray}
\frac{\partial^2 {\bar\Phi}_l}{\partial\tau^2} & - &
\frac{2}{\tau} \frac{\partial {\bar\Phi}_l}{\partial\tau} -
\frac{2}{r} \frac{\partial {\bar\Phi}_l}{\partial r} -
\frac{\partial^2 {\bar\Phi}_l}{\partial r^2} - \left[\frac{l\left(
l+1\right)}{r^2} - M^2 a^2(\tau)\right] {\bar\Phi_l}  \nonumber  \\
& = & \left( 1- \frac{J}{F}\right) \frac{\partial^2
{\bar\Phi}_l}{\partial\tau^2} - \left[ \frac{2}{\tau} +
\frac{1}{\sqrt{F J}} \frac{\partial}{\partial\tau}
\left(\frac{J^3}{F}\right)^{1/2} \right] \frac{\partial
{\bar\Phi}_l}{\partial\ \tau} \nonumber \\
& - & M^2 \left(J -1\right) {\bar\Phi}_l + \frac{1}{2} \left(
\frac{1}{F} \frac{\partial F}{\partial r} + \frac{1}{J}
\frac{\partial J}{\partial r}\right)
\frac{\partial\bar\Phi_l}{\partial r}. \label{a11}
\end{eqnarray}
Next,using the fact that $\epsilon$ is a small parameter, we propose the following expansion for
$\bar\Phi_l $ in orders of $\epsilon$:
\begin{equation}\label{a13}
\bar\Phi_l(r,\tau) = \bar\Phi^{(0)}_l + \bar\Phi^{(1)}_l +
\bar\Phi^{(2)}+ ....
\end{equation}
If we expand the right hand side of the equation (\ref{a11}) as
powers of $\epsilon \ll 1 $ \cite{CNW}, we obtain
\begin{eqnarray}
&- & 8 \left(\frac{\epsilon\tau}{2 r}\right) \left[
\frac{\partial^2{\bar\Phi}^{(0)}_l}{\partial\tau^2} -
\frac{1}{\tau} \frac{\partial {\bar\Phi}^{(0)}_l}{\partial\tau} -
\frac{M^2}{2H^2\tau^2} {\bar\Phi}^{(0)}_l\right] \nonumber \\
& - & 30 \left(\frac{\epsilon \tau}{2 r}\right)^2 \left[
\frac{\partial^2  {\bar\Phi}^{(1)}_l}{\partial \tau^2} -
\frac{1}{15} \frac{1}{\tau} \frac{\partial
{\bar\Phi}^{(1)}_l}{\partial {\bar\Phi}_l} - \frac{M^2}{5 H^2
\tau^2}{\bar\Phi}^{(1)}_l\right] + ... \label{a12}
\end{eqnarray}
Thus, the spectrum (\ref{fluct}) can be written using the expansion (\ref{a13}) as
\begin{eqnarray}
{\cal P}_k(\tau) & = & \sum_{l} \frac{(2l+1)}{4\pi} k^5 \,j^2_l(k
r) \left[ \bar\Phi^{(0)}_{kl} + \bar\Phi^{(1)}_{kl} + ...\right]
\left[ \left(\bar\Phi^{(0)}_{kl}\right)^* +
\left(\bar\Phi^{(1)}_{kl}\right)^* + ... \right] \nonumber \\
& = & \frac{k^3}{2\pi^2} \left|\bar\Phi^{(0)}_{kl=0}\right|^2 +
\frac{H^2}{4\pi^2} \epsilon \sum^{\infty}_{l=1}\, (2l+1)\, j^2_l(k
r)\, \Delta^{(1)}_{kl} + ...\, , \label{spect}
\end{eqnarray}
where
\begin{equation}\label{spect1}
\Delta^{(1)}_{kl} = \left(\frac{4\pi^2}{H^2 \epsilon}\right)
\frac{k^5}{4\pi} \left|\bar\Phi^{(0)}_{kl}
\left(\bar\Phi^{(1)}_{kl}\right)^* + \bar\Phi^{(1)}_{kl}
\left(\bar\Phi^{(0)}_{kl}\right)^*\right] = \frac{2\pi}{H^2
\epsilon} k^5 {\rm Re}
\left[\bar{\Phi}^{(1)}_{kl}\,\left(\bar\Phi^{(0)}_{kl}\right)^*\right].
\end{equation}
Notice that the first term in (\ref{spect}) corresponds to $l=0$,
so that the zeroth order approximation in $\epsilon$ is due only
to isotropic fluctuations. Terms with $l=1$ correspond to dipoles
and $l\geq 2$ are related to multipoles.

\section{Coarse-graining of $\bar{\varphi}$}

As it was shown in \cite{antigravity}, the metric (\ref{b1})
written in the static coordinate chart $(T,R)$, describes an
spherically symmetric object having properties of attractive and
repulsive gravity, under the election of $\psi_{0}=H^{-1}$.
Specifically, at scales larger than the
gravitational-antigravitational radius $R_{ga}$ , gravity
manifests itself as repulsive in nature. On the contrary, on
scales smaller than $R_{ga}$ gravity recovers its usual attractive
behavior. In this section our goal is to study the evolution of
the effective scalar field $\bar{\varphi}$ under the presence of
such an object but in the dynamical coordinate chart $(\tau,\vec{r})$.

To study the evolution of the effective field $\bar{\varphi}
(\tau,\vec{r})$ on scales larger than the
gravitational-antigravitational radius $r_{ga}$ we introduce the
field
\begin{equation}\label{c1}
\bar{\varphi}_{L}(\tau,\vec{r})=\int_{k_{H}}^{k_{Sch}}dk \sum_{l,m}\Theta_{L}(\sigma k_{ga}-k)
\left[a_{klm}\bar{\Phi}_{klm}(\tau,\vec{r})
+a_{klm}^{\dagger}\bar{\Phi}_{klm}^{*}(\tau,\vec{r})\right],
\end{equation}
where $\Theta _{L}$ is denoting the heaviside function, and the
wave number associated to the Hubble horizon
is
\begin{equation}\label{kh} k_{H}(\tau)\simeq
2\pi/[a(\tau)r_{H}]=-(2\pi)H\tau/r_{H}.
\end{equation}
Furthermore, the time dependent wavenumber
\begin{equation}\label{kga}
k_{ga}(\tau)=[2\pi/(a(\tau)r_{ga})][(2a(\tau)r_{ga})/(2a(\tau)r_{ga}+Gm)]^{2},
\end{equation}
is the wave number associated to the
gravitational-antigravitational radius $r_{ga}$, and $\sigma$ is a
dimensionless parameter that during inflation ranges in the
interval $10^{-3}-10^{-2}$.

Similarly, the evolution of the effective scalar field
$\bar\varphi (\tau,\vec{r})$ on small scales: scales between the
Schwarzschild radius $r_{Sch}$ and the
gravitational-antigravitational radius $r_{ga}$, can be described
by the field
\begin{equation}\label{c2}
\bar{\varphi}_{S}(\tau,\vec{r})=\int_{k_{H}}^{k_{Sch}} dk\sum_{l,m} \Theta_{S}(k-\sigma k_{ga})\left[a_{klm}\bar{\Phi}_{klm}
(\tau,\vec{r})+a_{klm}^{\dagger}\bar{\Phi}_{klm}^{*}(\tau,\vec{r})\right],
\end{equation}
where $\Theta_{S}$ denotes the heaviside function and $k_{Sch}\simeq 8\pi a(\tau)r_{Sch}/(Gm)^{2}=-8\pi r_{Sch}/[H\tau (Gm)^{2}]$
is the wave number associated to the Schwarzschild radius $r_{Sch}$. From the expressions (\ref{c1}) and (\ref{c2}) it can be easily
seen that $\bar\varphi(\tau,\vec{r})=\bar\varphi_{L}(\tau,\vec{r})+\bar\varphi_{S}(\tau,\vec{r})$.\\

\section{Scalar field fluctuations at zeroth order in $\epsilon$}

At zeroth order in the expansion (\ref{a13}), the equation (\ref{a11}) reduces to
\begin{equation}\label{z1}
\frac{\partial ^{2} \bar{\Phi}^{(0)}_l}{\partial
\tau^2}-\frac{2}{\tau}\frac{\partial \bar{\Phi}^{(0)}_l}{\partial
\tau}-\frac{2}{r} \frac{\partial\bar{\Phi}^{(0)}_l}{\partial
r}-\frac{\partial^{2}\bar{\Phi}^{(0)}_l}{\partial
r^2}-\left[\frac{l(l+1)}{r^2}-M^2
a^2(\tau)\right]\bar{\Phi}^{(0)}_l=0,
\end{equation}
where for the zeroth approximation we must restrict to  $l=0$. Now
in order to simplify the structure of (\ref{z1}), let us to
introduce the field $\chi^{(0)}_{l=0}(\tau,r)$, with
$\bar{\Phi}^{(0)}_{l=0}(\tau,r)=\tau\chi^{(0)}_{l=0}(\tau,r)$, so
that the equation (\ref{z1}) can be written in the form
\begin{equation}\label{z3}
\frac{\partial^{2}\chi^{(0)}_{l=0}}{\partial\tau
^2}-\frac{2}{r}\frac{\partial\chi^{(0)}_{l=0}}{\partial
r}-\frac{\partial ^{2}\chi^{(0)}_{l=0}}{\partial r^2}
-m_{eff}^{2}(\tau)\,\chi^{(0)}_{l=0}=0,
\end{equation}
where $m^{2}_{eff}(\tau)=2/\tau^2-M^2/(H^2 \tau^2)$ is the
effective mass of the inflaton field. By means of the Bessel
transformation
\begin{equation}\label{z4}
\chi^{(0)}_{l=0}(\tau,r)=\int_{0}^{\infty}dk k^2
j_{{l=0}}(kr)\xi_{k {l=0}}^{(0)}(\tau)
\end{equation}
we derive from (\ref{z3}) the next equation for the modes
$\xi_{k0}$:
\begin{equation}\label{z5}
\frac{\partial^{2}\xi_{k0}^{(0)}}{\partial
\tau^2}+\left[k^{2}-m_{eff}^{2}(\tau)\right]\xi_{k0}^{(0)}=0,
\end{equation}
such that the modes of $\bar{\Phi}_{{l=0}}^{(0)}$ are given by
$\bar{\Phi}_{k0}^{(0)}=\tau \xi_{k0}^{(0)}$. Thus solving
(\ref{z5}) the normalized solution for the modes
$\bar{\Phi}_{k0}^{(0)}$ has the form
\begin{equation}\label{zz5}
\bar{\Phi}_{k0}^{(0)}(\tau)= A_1 \, \left(-\tau\right)^{3/2} {\cal
H}^{(1)}_{\nu}\left[-k\,\tau\right] + A_2 \,
\left(-\tau\right)^{3/2} {\cal
H}^{(2)}_{\nu}\left[-k\,\tau\right],
\end{equation}
Here, ${\cal H}^{(1,2)}_{\nu}[-k\tau]$ are respectively the first
and second kind Hankel functions, $\nu^2= {9\over 4} - {M^2\over
H^2}$, and the normalization constants are given by
\begin{equation}\label{zz51}
A_2 = -\frac{\sqrt{\pi} H}{2}\,e^{-i \nu \pi/2}, \qquad A_1 = 0.
\end{equation}
Now we introduce the fields
\begin{eqnarray}\label{z6}
\left[\chi^{(0)}_{L}\right]_{l=0}(\tau,r)&=&
\int_{k_{H}}^{k_{Sch}}dk\, \Theta_{L}(\sigma
k_{ga}-k)\left[a_{k0}j_{0}(kr)\xi_{k0}^{(0)}(\tau)+
a_{k0}^{\dagger}j_{{0}}^{*}(kr)\xi_{k0}^{(0)}\,^{*}(\tau)\right],\\
\label{z7} \left[\chi^{(0)}_{S}\right]_{l=0}(\tau,r)&=&
\int_{k_{H}}^{k_{Sch}}dk\, \Theta_{S}(k-\sigma
k_{ga})\left[a_{k0}j_{0}(kr)\xi_{k0}^{(0)}(\tau)+
a_{k0}^{\dagger}j_{0}^{*}(kr)\xi_{k0}^{(0)}\,^{*}(\tau)\right],
\end{eqnarray}
where
$\chi^{(0)}_{l=0}(\tau,r)=\left[\chi^{(0)}_{L}\right]_{l=0}(\tau,r)+\left[\chi_{S}^{(0)}\right]_{l=0}(\tau,r)$
and $\xi_{k0}^{(0)}(\tau)=\tau^{-1}\bar{\Phi}_{k0}^{(0)}(\tau)$.
The equation of motion for $\left[\chi_{L}^{(0)}\right]_{l=0}$ is
given by
\begin{equation}\label{z8}
\left[\ddot{\chi}_{L}^{(0)}\right]_{l=0}-m_{eff}^{2}(\tau)\left[\chi_{L}^{(0)}\right]_{l=0}=\sigma\ddot{k}_{ga}\eta^{(0)}_{l=0}(\tau,r)
+\sigma\dot{k}_{ga}\lambda^{(0)}_{l=0}(\tau,r)
+2\sigma\dot{k}_{ga}\gamma^{(0)}_{l=0}(\tau,r) ,
\end{equation}
where the stochastic operator fields $\eta^{(0)}_{l=0}$,
$\lambda^{(0)}_{l=0}$ and $\gamma^{(0)}_{l=0}$ are defined as
\begin{eqnarray}\label{z9}
\eta^{(0)}_{l=0}(\tau,r)=\int_{k_H}^{k_{Sch}}dk\, \delta(k-\sigma
k_{ga})\left[a_{k0}j_{0}(kr)\xi_{k0}^{(0)}(\tau)+a_{k0}^{\dagger}j_{0}^{*}(kr)
\xi_{k0}^{(0)}\,^{*}(\tau)\right],\\
\label{z10} \lambda^{(0)}_{l=0}(\tau,r)=\int_{k_H}^{k_{Sch}}dk\,
\dot{\delta}(k-\sigma
k_{ga})\left[a_{k0}j_{0}(kr)\xi_{k0}^{(0)}(\tau)+a_{k0}^{\dagger}j_{0}^{*}(kr)
\xi_{kl}^{(0)}\,^{*}(\tau)\right],\\
\label{z11} \gamma^{(0)}_{l=0}(\tau,r)=\int_{k_H}^{k_{Sch}}dk\,
\delta(k-\sigma
k_{ga})\left[a_{k0}j_{0}(kr)\dot{\xi}_{k0}^{(0)}(\tau)+a_{k0}^{\dagger}j_{0}^{*}(kr)
\dot{\xi}_{k0}^{(0)}\,^{*}(\tau)\right],
\end{eqnarray}
with the dot denoting $\partial/\partial \tau$. The field equation (\ref{z8}) can be expressed in the form
\begin{equation}\label{z12}
\left[\ddot{\chi}_{L}^{(0)}\right]_{l=0}-m_{eff}^{2}(\tau)\left[\chi_{L}^{(0)}\right]_{l=0}=\sigma\left[\frac{d}{d\tau}(\dot{k}_{ga}\eta^{(0)}_{l=0})
+2\dot{k}_{ga}\gamma^{(0)}_{l=0}\right].
\end{equation}
This is a Kramers-like stochastic equation, that with the help of
the auxiliary field:
$u^{(0)}_{l=0}=\left[\dot{\chi}_{L}^{(0)}\right]_{l=0}-\sigma\dot{k}_{ga}\eta
^{(0)}_{l=0}$, can be written as the first order stochastic system
\begin{eqnarray}\label{z13}
\dot{u}^{(0)}_{l=0}&=&
m_{eff}^{2}\left[\chi_{L}^{(0)}\right]_{l=0}
+2\sigma\dot{k}_{ga}\gamma^{(0)}_{l=0},\\
\label{z14} \left[\dot{\chi}_{L}^{(0)}\right]_{l=0}&=&
u^{(0)}_{l=0}+\sigma\dot{k}_{ga}\eta^{(0)}_{l=0},
\end{eqnarray}
The role that the noise $\gamma^{(0)}_{l=0}$ plays in this system,
can be minimized  in the system (\ref{z13}) and (\ref{z14}) when
the condition
$\dot{k}_{ga}^{2}\left<(\gamma^{(0)}_{l=0})^2\right>\ll
\ddot{k}_{ga}^{2}\left<(\eta^{(0)}_{l=0})^{2}\right>$ is valid.
This condition can be expressed as
\begin{equation}\label{z15}
\left.\frac{\dot{\xi}_{ k0}^{(0)}(\dot{\xi}_{
kl}^{(0)})^{*}}{\xi_{k0}^{(0)}(\xi_{k0}^{(0)})^{*}}\right|_{k=\sigma
k_{ga}}\ll\, \left(\frac{\ddot{k}_{ga}}{\dot{k}_{ga}}\right)^{2},
\end{equation}
which is valid on large scales i.e. scales where $
k_{ga}(\tau)<k<k_{H}(\tau)$, for $k_H$ and $k_{ga}$ given
respectively by (\ref{kh}) and (\ref{kga})\footnote{When the
background is an exact de Sitter space-time and the field is free,
this condition is analogous to one obtained already by
Mijic\cite{Mij} in a different approach.}. If this is the case,
the system (\ref{z13}) and (\ref{z14}) can be approximated by
\begin{eqnarray}\label{z16}
\dot{u}^{(0)}_{l=0}&=& m_{eff}^{2}\left[\chi_{L}^{(0)}\right]_{l=0},\\
\label{z17} \left[\dot{\chi}_{L}^{(0)}\right]_{l=0}&=&
u^{(0)}_{l=0}+\sigma\dot{k}_{ga}\eta^{(0)}_{l=0}.
\end{eqnarray}
This is an stochastic two-dimensional Langevin equation with a
noise $\eta^{(0)}_{l=0}$ which is gaussian and white in nature, as
it is indicated by the following expressions:
\begin{eqnarray}\label{z18}
\left<\eta^{(0)}_{l=0}\right>&=& 0,\\
\label{z19} \left<(\eta^{(0)}_{l=0})^2\right>&=& 4\pi\sigma
\frac{k_{ga}^{2}}{\dot{k}_{ga}}\left.j_{0}(kr)j_{0}^{*}(kr)\xi_{k0}^{(0)}\xi_{k0}^{(0)}\,^{*}
\right|_{k=\sigma k_{ga}}\delta(\tau-\tau^{\prime}).
\end{eqnarray}
The correlation functions of $\eta^{(0)}_{l=0}$ and
$\gamma^{(0)}_{l=0}$ have the same structure, similar to the
momenta of a Gaussian white noise. The dynamics of the probability
transition ${\cal
P}^{(0)}_{l=0}\left[\left[\chi_{L}^{I}\,^{(0)}\right]_{l=0},\left[u^{I}\,^{(0)}\right]_{l=0}|\left[\chi_{L}^{(0)}\right]_{l=0},u^{(0)}_{l=0}\right]$
from an initial configuration
$(\left[\chi_{L}^{I}\,^{(0)}\right]_{l=0},\left[u^{I}\,^{(0)}\right]_{l=0})$
to a configuration $\left(\chi_{L}^{(0)},u^{(0)}\right)$, is given
by the Fokker-Planck equation:
\begin{equation}\label{z20}
\frac{\partial {\cal P}^{(0)}_{l=0}}{\partial
\tau}=-u^{(0)}\frac{\partial{\cal
P}^{(0)}_{l=0}}{\partial\left[\chi_{L}^{(0)}\right]_{l=0}}-m_{eff}^{2}\left[\chi_{L}^{(0)}\right]_{l=0}
\frac{\partial {\cal P}^{(0)}_{l=0}}{\partial
u^{(0)}_{l=0}}+\frac{1}{2}D_{11}^{(0)}\frac{\partial^{2}{\cal
P}^{(0)}_{l=0}}{\partial \left[\chi_{L}^{(0)}\right]^{2}_{l=0}},
\end{equation}
where $D_{11}^{(0)}=\int
(\sigma\dot{k}_{ga})^{2}\left<(\eta^{(0)}_{l=0})^2\right> d\tau$
is the diffusion coefficient related to
$\left[\chi_{L}^{(0)}\right]_{l=0}$. By using (\ref{z19}) the
diffusion coefficient $D_{11}^{(0)}$ becomes
\begin{equation}\label{z21}
D_{11}^{(0)}=4\pi \sigma^{3}\dot{k}_{ga}k_{ga}^{2}\left.
j_{0}(kr)j_{0}^{*}(kr)\xi_{k0}^{(0)}\xi_{k0}^{(0)}\,^{*}\right|_{k=\sigma
k_{ga}}.
\end{equation}
Hence, the dynamics of
$\left<\left(\left[\chi_{L}^{(0)}\right]_{l=0}\right)^2\right>=\int
d\left[\chi_{L}^{(0)}\right]_{l=0}\, du^{(0)}_{l=0}
\left(\left[\chi_{L}^{(0)}\right]_{l=0}\right)^{2}{\cal
P}^{(0)}_{l=0}$ is given by the equation
\begin{equation}\label{z22}
\frac{d}{d\tau}\left<\left(\left[\chi_{L}^{(0)}\right]_{l=0}\right)^2\right>
= \frac{1}{2}D_{11}^{(0)}(\tau).
\end{equation}
Now, in order to return to the original zeroth order scalar field,
let us to use the expression
$\bar{\Phi}^{(0)}_{l=0}(\tau,r)=\tau\chi^{(0)}_{l=0}(\tau,r)$  in
(\ref{z22}) to obtain
\begin{equation}\label{z23}
\frac{d}{d\tau}\left<\left(\left[\bar{\phi}_{L}^{(0)}\right]_{l=0}\right)^2\right>
=
\frac{2}{\tau}\left<\left(\left[\bar{\phi}_{L}^{(0)}\right]_{l=0}\right)^2\right>
+ \frac{1}{2}\tau^{2}D_{11}^{(0)}(\tau) .
\end{equation}
The general solution of (\ref{z23}), is then
\begin{equation}\label{z24}
\left<\left(\left[\bar{\phi}_{L}^{(0)}\right]_{l=0}\right)^2\right>=\frac{1}{2}\tau^{2}\left[\int^{\tau}
D_{11}^{(0)}(\tau') d\tau' + C\right],
\end{equation}
with $C$ an integration constant. Next, we employ the relation
\begin{equation}\label{z25}
\left<\left(\bar{\varphi}^{(0)}_{L}\right)^2\right>=
\left(\frac{1}{4\pi}\right)\left<\left(\left[\bar{\Phi}_{L}^{(0)}\right]_{l=0}\right)^2\right>,
\end{equation}
where we have used the addition theorem of the spherical harmonics, to derive the equation
\begin{equation}\label{z26}
\left<(\bar{\varphi}^{(0)}_{L})^2\right>=\left(\frac{1}{8\pi}\right)\tau^{2}\left[\int
D_{11}^{(0)}(\tau) d\tau + C\right].
\end{equation}
This equation, give us in principle the squared fluctuations of
$\bar{\varphi}_{L}$ on large scales. Employing (\ref{zz5}) and
(\ref{z21}), the expression (\ref{z26}) with $C=0$, can be
approximated on the IR sector as
\begin{equation}
\left.\left<\left(\bar{\varphi}_{L}^{(0)}\right)^{2}\right>\right|_{kr
\ll 1}  \simeq  \,\left(\frac{H^2}{2\pi}\right)
\,2^{2(\nu-1)}\sigma^{3-2\nu} \Gamma^{2}(\nu)\,
(-\tau)^{3-2\nu}\int\frac{dk_{ga}}{k_{ga}}k_{ga}^{3-2\nu}\,
\label{z27}
\end{equation}
where we have used the asymptotic expansion $j_{0}(kr)|_{kr\ll
1}\simeq 1$. The spectrum derived from (\ref{z27}) at zeroth order
(i.e. for $l=0$), has the form
\begin{equation}\label{z27aa}
{\cal P}_{k_{ga}}^{(0)}(\tau)\simeq 2^{2(\nu-1)} \,\Gamma^{2}(\nu)
\left(\frac{H^2}{2\pi}\right) \left[\sigma (-\tau)
k_{ga}\right]^{3-2\nu},
\end{equation}
 which results scale invariant when $\nu=3/2$ and when this is the case its time dependence disappears.
 This spectrum is similar to whole obtained in a previous
 work\cite{ultimo}, but in a different manner. It is
 characteristic of an universe governed by a cosmological constant, in agreement with one expects for a de Sitter expansion
of the universe. However, for spectrums close, but different to
the Harrison-Zeldovich, the situation changes, because it becomes
sensitive to the wavenumber $k_{ga}$ and therefore with the mass
of the BH. Furthermore, the amplitude of this spectrum tends to
zero (as $\tau \rightarrow 0$), for $\nu < 3/2$.

\section{Final Comments}

We have developed a stochastic approach to study scalar field
fluctuations of the inflaton field in an early inflationary
universe, which is described by an effective 4D SdS metric. The
cosmological metric was obtained using planar coordinate
transformations on a 5D Ricci-flat Schwarzschild-de Sitter (SdS)
static metric (\ref{a1}), for a SdS BH. From the dynamical point
of view, the effective 4D cosmological metric (\ref{b1}) describes
the collapse of the universe on scales $k \gg k_{ga}$ and an
accelerated expansion for scales much bigger than the
gravitational - antigravitational radius $r_{ga}$, which is
related with the wavenumber $k_{ga}$.

The main difference with earlier stochastic approaches to
inflation where the window function is defined on the Hubble
horizon is that, in our approach [see eq. (\ref{c1})], the
coarse-grained field is defined using a window function
$\Theta_L(\sigma k_{ga} - k)$, which takes into account only modes
with wavelengths larger than the gravitational - antigravitational
radius $r_{ga}$. This fact indicates the scale for which the
universe is starting to expand accelerated. On smaller scales the
universe is collapsing due to the attraction of the BH. However,
on larger scales gravitation is repulsive and drives inflation.
For the limit case in which this mass is very small, $G m/(2\, a
r_{ga}) \ll 1$, we obtain that $\left.k_{ga}\right|_{G m /(2 \,a
r_{ga}) \ll 1} \rightarrow k_{H}$, and our result agrees
completely with whole of the squared field fluctuations of a de
Sitter expansion during the inflationary stage when the horizon
entry. For $r\rightarrow \infty$ $J$ and $F$ approach to
$a^2(\tau)$ and the metric (\ref{b1}) describes a de Sitter
expansion. However, for very large (but finite) cosmological
scales the spectrum is not exactly scale independent, because
becomes sensitive to the wavenumber $k_{ga}$. For $\nu < 3/2$ the
spectral index $n_s = 3-2\nu$ is positive and the amplitude
decreases as $\tau \rightarrow 0$.

We have restricted our stochastic study to very small fluctuations
on cosmological scales. A more profound study should necessarily
include higher orders in the expansion (\ref{a13}) of $\bar\phi_l$
in the equation (\ref{a11}), which takes into account multipolar
expansion due to non-gaussian noises.

\section*{Acknowledgements}

\noindent J.E.M.A and L.M. Reyes acknowledge CONACYT (M\'exico) and M.B.
acknowledges UNMdP and CONICET (Argentina) for financial support.

\end{document}